\documentclass[]{llncs}

\usepackage{amsmath}
\usepackage{amssymb}
\usepackage{graphicx}
\usepackage{subcaption}

\begin{document}

\title{Compressed Game Solving}

\author{Jeffrey Considine\orcidID{0009-0004-5587-5979}}
\institute{Boston University \\ \email{jconsidi@bu.edu}}

\date{}

\maketitle

\begin{abstract}
  We recast move generators for solving board games as operations on compressed sets of strings.
  We aim for compressed representations with space sublinear in the number of game positions for interesting sets of positions, move generation in time roughly linear in the compressed size and membership tests in constant time.
  To the extent that we achieve these tradeoffs empirically, we can strongly solve board games in time sublinear in the state space.
  We demonstrate this concept with the game Breakthrough where we empirically realize compressed representations taking roughly $n^{0.5}$ to $n^{0.7}$ space to store relevant sets of $n$ positions.
\end{abstract}

\section{Introduction}

Computer game playing has been an interest nearly as long as general purpose computers have existed; Alan Turing's ``Proposed Electronic Calculator'' report in 1946 predicted that computers ``could probably be made to play very good chess''~\cite{turing1946proposed}.
The idea of methodically solving a game to determine the winner under perfect play goes back farther to Zermelo's theorem in 1913~\cite{schwalbe2001zermelo}; previously, it was not even clear that games could be solved in general.
The first example of solving a game is even older -- the game of Nim was strongly solved in~\cite{bouton1901nim}, and there are likely older examples of small games such as Tic-tac-toe being solved.
So what keeps us from solving all the games of interest?

Interesting games tend to have too many positions for our generic techniques to work, so we end up looking for knowledge-based short cuts or investing inordinate amounts of compute power.
If the game is too small or a trick is too powerful, we lose interest in the game.
For example, Tic-tac-toe is often solved informally by elementary school children, and Nim has a trivial to calculate rule to determine both the winner and ideal moves.
Early game solutions such as Qubic~\cite{patashnik1980qubic}, Connect 4~\cite{allis1988knowledgebased}, and Gomoku~\cite{allis1993gomoku} were achieved by integrating knowledge into the solving program to significantly reduce the search space.

The earliest non-trivial game solved without a substantial advantage from knowledge is generally held to be Nine Men's Morris which solved by Gasser in 1993~\cite{gasser1996solving}.
The solution of Nine Men's Morris comprised of an endgame database of about $10^{10}$ states solving the midgame and endgame phases and an 18 ply alphabeta search from the beginning of the game to the midgame.
At the time, this was a non-trivial amount of resources.
Since this first interesting solution, Checkers has also been shown to be a draw through a combination of a ten piece endgame database, and proof number search from the beginning of the game~\cite{schaeffer2007checkers}.
More recently, Othello was also shown to be drawn~\cite{takizawa2023othello}.
Table~\ref{table:solved-games} provides a longer list of games solved with such strategies.

\begin{table}[ht]
  \begin{center}
    \begin{tabular}{|c|c|c|c|c|}
      \hline
      Game & Year & State Space & Positions Solved & Solve Strength \\
      \hline
      \hline
      Nine Men's Morris & 1993~\cite{gasser1996solving} & $10^{10}$ & $10^{10}$ & strong \\
      Awari & 2002~\cite{romein2002awari} & $9 \times 10^{11}$ & $9 \times 10^{11}$ & strong \\
      Checkers & 2007~\cite{schaeffer2007checkers} & $5 \times 10^{20}$ & $3.9 \times 10^{13}$ & weak \\
      Fanorona & 2008~\cite{schadd2008best} & $10^{21}$ & $6.3 \times 10^{9}$ & weak \\
      Pentago & 2014~\cite{irving2014pentago} & $3 \times 10^{15}$ & $3 \times 10^{15}$ & strong \\
      Othello & 2023~\cite{takizawa2023othello} & $10^{28}$ & $1.5 \times 10^9$ & weak \\
      \hline
    \end{tabular}
  \end{center}

  \caption{Games solved with the help of endgame databases.}
  \label{table:solved-games}
\end{table}

A strategy shared across these solutions was the identification of a set of intermediate positions whose solution would prune a substantial fraction of a search tree from the beginning of the game.
That intermediate set of positions was then solved via a brute force method (retrograde analysis except for Othello), and then a search from the root was used to construct a proof solving the starting position of the game.
A key design decision is identifying the set of intermediate positions to solve.
A smaller set of intermediate positions will require a larger search process, while a larger set of intermediate positions will require more time to solve.
In previous work, solving the set of intermediate positions took time at least linear in the size of that set; linear time was required just to write the solutions, and individual positions might need to be processed multiple times to resolve them.
The most common strategy, retrograde analysis, solves a set of intermediate positions that is closed under reachability by working from the end of the game, and efficient implementations can be close to that linear ideal.
Our work aims to break that linear time requirement by compressing the sets of positions into sublinear representations, and performing move generation directly on those compressed sublinear representations.

Our contributions are as follows.
We recast the move generation operations of retrograde analysis as set operations, similar to the exposition by Von Neumann and Morgenstern~\cite{vonneumann1945theory}, and argue that suitable compressed set representations can radically change the cost of retrograde analysis.
We then show how to instantiate this compressed set approach using deterministic finite automata.
We demonstrate this approach for multiple games showing its strengths and weaknesses, and ultimately solve several new sizes of the game Breakthrough on a single commodity laptop.
We also show that this particular implementation has more modest leverage for the games of Amazons and Chess, and close speculating on possible improvements to the compressed set representations.

\section{Retrograde Analysis using Sets}

The most common strategy for solving interesting games has been retrograde analysis.
Retrograde analysis starts at the end of the game with positions defined as won or lost, and iteratively solves positions that require one more ply (turn) before the winning side can force its win.
We will now sketch this process, roughly following Von Neumann and Morgenstern~\cite{vonneumann1945theory} due to their focus on sets, but focusing more on cumulatively solved sets without their series of recursively defined games.

Let $P$ be the set of all positions of the game and $T$ be the set of terminal positions where the game has ended.
Let $W_0 \subseteq T$ be the set of positions where the game has ended and the current player has won, and $L_0 \subseteq T$ be the set of positions where the game has ended and the current player has lost.
Both $W_0$ and $L_0$ will be provided as part of the game definition.
For $i > 0$, let $W_i$ and $L_i$ be the sets of positions where a player can force a win within $i$ ply, and the current player respectively will win or lose.
Then,

\begin{eqnarray*}
  W_{i+1} & = & W_0 \cup \mathrm{reverse}(L_{i}) \\
  L_{i+1} & = & L_0 \cup \left( \mathrm{inverse}(T) \setminus \mathrm{reverse}(\mathrm{inverse}(W_{i})) \right) \\
\end{eqnarray*}

\noindent
where

\begin{eqnarray*}
  \mathrm{inverse}(S) & = & P \setminus S \\
  \mathrm{reverse}(S) & = & \{ p \in P~|~\exists p_S \in S~\mathrm{s.t.~there~is~a~move~from}~p~\mathrm{to}~p_S \} \\
\end{eqnarray*}

\noindent
We can completely solve the game if we compute
\begin{eqnarray*}
  W_\infty & = & \lim_{i \rightarrow \infty} W_i \\
  L_\infty & = & \lim_{i \rightarrow \infty} L_i \\
\end{eqnarray*}

In the terminology of Allis, constructing $W_\infty$ and $L_\infty$ is sufficent to strongly solve the game~\cite{allis1994searching}, since we can lookup any position in each set immediately, and if it is in neither, the position is drawn.
As long as $P$ is finite, $W_\infty = W_i$ and $L_\infty = L_i$ for some $i \leq |P|$~\cite{schwalbe2001zermelo,vonneumann1945theory}, though that $i$ may be exponentially larger than a natural encoding of a position~\cite{fraenkel1981computing}.

In previous practice, retrograde analysis worked with individual positions instead of sets.
Each time a new losing position $p_L$ is identified, a reverse move generator is used to identify positions $p$ that can move to $p_L$, and flag those positions $p$ as winning positions.
Each time a new winning position $p_W$ is identified, a reverse move generator to identify positions $p$ that can move to $p_W$, and if $p$ can only move to positions known to be winning, then flag $p$ as losing.
Altogether, the total work is linear in the number of positions solved, though there are many systems issues to manage to actually achieve that, particularly when the sets of winning and losing positions do not fit in memory and to avoid repeatedly calling the reverse move generator on the same position.
See~\cite{schaeffer2003building} for details.

In contrast to those previous methods, our approach will be focused on set operations, and will not work with individual positions.
Before proceeding, we note that based on the formulas above, our implementation will need to support set union ($\cup$), set difference ($\setminus$), and the $\mathrm{reverse}$ function.
The performance of these operations will determine the efficacy of our approach.

\section{Meet-in-the Middle Analysis Using Sets}

Many previous game solutions using retrograde analysis did not solve all of $P$.
Instead, these ``weak'' solutions~\cite{allis1994searching} only target solving the initial position of the game and any positions necessary to support that solution.
While small games such as $3 \times 3$ Tic-Tac-Toe can be solved with pure search, larger games such as Nine Men's Morris~\cite{gasser1996solving}, Checkers~\cite{schaeffer2007checkers}, and Fanorama~\cite{schadd2008best} used a meet-in-the-middle approach using more sophisticated proof search techniques pruned when they reached positions solved by retrograde analysis.
For ``converging'' games~\cite{allis1994searching}, such as when captures irreversibly decrease the number of pieces on the board, this strategy can be particularly effective.
Such proof search techniques work just as well whether the retrograde analysis was computed with individual positions or sets.
It is also possible to supplement or replace these proof search techniques with a set-based approach which we will describe now.

In contrast to proof search techniques which rely heavily on pruning, our set-based approach to meet-in-the-middle analysis starts with constructing sets reachable at a given number of ply from the beginning of the game.

\begin{eqnarray*}
  R_0 & = & \{ p \in P~|~\mathrm{s.t.}~p~\mathrm{is~the~initial~position} \} \\
  R_{i + 1} & = & \mathrm{forward}(R_i) \\
\end{eqnarray*}

Define $RW_{i,j}$ to be the set of positions reachable in $i$ ply that will win within another $j$ ply, and $RL_{i,j}$ to be the set of positions reachable in $i$ ply that will lose within another $j$ ply.
Then define $RU_{i,j}$ as the set of positions reachable in $i$ ply that do not have a win or loss forced within another $j$ ply.
By those definitions,

\begin{eqnarray*}
  RW_{i,j} & = & R_{i} \cap W_j \\
  RL_{i,j} & = & R_{i} \cap L_j \\
  RU_{i,j} & = & R_{i} \setminus (RW_{i,j} \cup RL_{i,j}) \\
\end{eqnarray*}

\noindent
We can also write recursive versions of these equations.

\begin{eqnarray*}
  RW_{i,j+1} & = & R_{i} \cap (W_0 \cup \mathrm{reverse}(RL_{i+1,j})) \\
  RL_{i,j+1} & = & R_{i} \cap (L_0 \cup (R_i \setminus \mathrm{reverse}(\mathrm{inverse}(RW_{i+1,j})))) \\
\end{eqnarray*}

\noindent
There, the recursive equations tradeoff the forward ply $i$ and backward ply $j+1$ to forward ply $i+1$ and backward ply $j$.
That recursion can be stopped at whatever number of backward ply $j$ has $W_j$ and $L_j$ computed via retrograde analysis.
And if $RU_{i+1,j} = \emptyset$, then all of $R_{i+1}$ was solved within an additional $j$ ply, so all of $R_i$ will be solved with an additional $j+1$ ply and we can compute $RL_{i,j+1}$ more simply via set difference.

\begin{eqnarray*}
  RL_{i,j+1} & = & R_i \setminus RW_{i,j+1} \\
\end{eqnarray*}

\noindent
Thus, if for some $i' > 0$ we can completely solve $R_{i'}$, then we can back up the solution to the initial position in $R_0$ with just $i'$ invocations of $\mathrm{reverse}$, set union, set intersection, and set difference.
Empirically, the main gain there is from dropping from $2 i'$ to $i'$ invocations of $\mathrm{reverse}$, since each invocation of $\mathrm{reverse}$ contains many similiarly sized invocations of the set operations.
For games with bounded numbers of moves, this also appears favorable compared to just retrograde analysis, since only one invocation of $\mathrm{reverse}$ is needed per ply vs 2-4 invocations for retrograde analysis depending on symmetries.
However, this is not necessarily always the case since retrograde analysis may need fewer ply to solve all positions that we care about, and intersections with reachable sets of positions may increase the complexity of the sets involved.

Most of the results that we present later will be computed with this meet in the middle approach using various numbers of ply for the pure retrograde analysis.
If we start our computations from an $i,j$ choice where $RU_{i,j}$ is empty, then we will solve all positions reachable within $i$ ply.
In that case, $RW_{i,j} = R_i \cap W_\infty$ and $RL_{i,j} = L_\infty$ and we denote them as $RW_{i,\infty}$ and $RL_{i,\infty}$ respectively.
This may not be a complete solution since positions that are at least $i+1$ ply from the beginning and needing at least $j+1$ additional ply to solve will be missed.
However, we expect that most of the missed positions will be easily solved in practice using search and the already computed $W_j$ and $L_j$.

\section{Compressed Game Solving}
\label{section:methods}

We have argued so far that a set-oriented perspective on game solving gives a simple view of the game solving process, but these arguments are moot if we cannot make use of a more efficient set representation.
We seek compressed set representations decreasing space usage by a fractional polynomial (e.g. $O(n^{0.5})$ space to store $n$ positions), time linear in that representation size for move generation, and very fast membership testing.
To be clear, we do not expect sublinear space usage for arbitrary sets of positions which would be impossible, but we hope for fractional polynomial space usage on interesting sets of positions such as ``first player to move and win within 20 ply''.
Using deterministic finite automata to represent positions, we empirically achieve sublinear space usage, heuristically linear move generation, and constant time membership testing.
The most similar previous application of compression used ordered binary decision diagrams to compress small chess endgame databases~\cite{kristensen2005generation}, but yielded much lower compression rates, and crucially, only compressed after solving all the positions.

Table~\ref{table:tradeoffs} compares the empirical tradeoffs that we seek versus previous approaches.
summarizes the tradeoffs that we aim to improve.
Specifically, it compares the space used by the representation, the time used for reverse move generaton, and the time used for membership tests to check if a position is in a set.
Our baseline, retrograde analysis, uses linear space in the number of the positions, linear time for move generation, and constant time for membership tests.
Previous compression work only compressed the output of retrograde analysis yielding constant factor improvements to space while preserving constant time membership tests~\cite{kristensen2005generation,gomboc2021chess}.
An alternative symbolic approach using quantified boolean formulas describes sets of positions with formulas taking time and space linear in the number of plys to construct those formulas, but taking exponential time in the number of ply to test membership.\footnote{The time to solve these QBF formulas is not actually proven to be exponential. Such a proof would imply PSPACE=EXPTIME since QBF satisfiability is in PSPACE.}
Or more extreme, the minimax algorithm would take space logarithmic in the number of ply (to record that number), and time exponential in the number of ply for membership testing.

\begin{table}[ht]
  \begin{center}
    \begin{tabular}{|l|c|c|c|}
      \hline
      Method & Representation Space & Move Generation Time & Membership Test Time \\
      \hline
      \hline
      Retrograde Analysis~\cite{vonneumann1945theory,strohlein1970untersuchungen,thompson1986retrograde} & $\Theta(n)$ & $\Theta(n)$ & $O(1)$ \\
      Retrograde Analysis (compressed output)~\cite{kristensen2005generation,gomboc2021chess} & $\Theta(n)$ & $\Theta(n)$ & $O(1)$ \\
      Compressed with deterministic finite automata & $O(n^c)$ (empirical) & $O(n^c)$ (empirical) & $O(1)$ \\
      Quantified boolean formulas~\cite{shaik2023implicit} & $\Theta(p)$ & $\Theta(p)$ & $2^{O(p)}$ \\
      Minimax (just ply saved) & $\Theta(\log p)$ & $\Theta(\log p)$ & $2^{O(p)}$ \\
      \hline
    \end{tabular}

    The costs in this table assume $n$ positions or $p$ ply.
  \end{center}

  \caption{Tradeoffs Solving Endgame Positions}
  \label{table:tradeoffs}
\end{table}

\subsection{Representing Positions as Strings}
\label{section:position-representation}

We represent game positions simply with a one-to-one mapping to the natural game state.
Most of the games that we consider are played on a rectangular board with pieces played on a grid of squares.
We represent positions from those games using a row-wise traversal with one character specifying the contents of each square.
We are aware that there are potential optimizations in changing the traversal order~\cite{kristensen2005generation}, but this order suffices to demonstrate our approach.

For the DFA alphabet, we use the first character for an empty board square for convenience sharing counting DFA code, and then enumerate the other possible contents of a board square as the remaining characters.
For the game Amazons, the alphabet is \texttt{empty}, \texttt{player 1 queen}, \texttt{player 2 queen}, and \texttt{arrow}.
For the game Breakthrough, the alphabet is \texttt{empty}, \texttt{player 1 piece}, \texttt{player 2 piece}.
For the game Chess, the alphabet is \texttt{empty}, black and white versions of each piece type, plus extra versions of pawns and rooks to represent en-passant and castling status.
This suffices for any game where all information of the position is visible looking at the board.
Other approaches are possible - extra characters could have been added at either end of the position string to denote the status for these special moves, but we felt this was the simplest mapping that would add the fewest complications to move generation.

\subsection{Generating Moves from Sets of Positions}

Generating moves from a set of positions mostly consists of set operations, but we will need to add an operation to apply particular moves to sets of positions.
First, let us define forward move generation.

\begin{eqnarray*}
  \mathrm{forward}(S) & = & \{ p \in P~|~\exists p_S \in S~\mathrm{s.t.~there~is~a~move~from}~p_S~\mathrm{to}~p \} \\
\end{eqnarray*}

\noindent
This definition is essentially the same as the $\mathrm{reverse}$ function used for retrograde analysis, but with the opposite direction in the move condition.
For any particular game, we will decompose $\mathrm{forward}$ into further operations to implement distinct moves.
As part of the definition of a game, we will require a list of moves $M_i$ where

\begin{eqnarray*}
  M_i & = & \langle \mathrm{Pre}_i, \mathrm{Changes}_i, \mathrm{Post}_i \rangle \\
  \mathrm{Changes_i} & = & [ \mathrm{Change}_{i,j} ] \\
  \mathrm{Change}_{i,j} & = & \langle \mathrm{Index}_{i,j}, \mathrm{Before}_{i,j}, \mathrm{After}_{i,j} \rangle \\
\end{eqnarray*}

\noindent
In essence, each move has a pre-condition, list of changes to the board, and a post-condition.
Both the pre- and post-conditions can be represented as sets of positions.
Each move's list of changes specifies each index in the string (position on the board) to change, its value before the move, and its value after the move.
To avoid ambiguity, the before and after values should be included in the pre- and post-conditions respectively.
With this specification, we can define a function $\mathrm{change}_i$ which applies the changes of move $M_i$ assuming the preconditions $\mathrm{Pre}_i$ hold.

\begin{eqnarray*}
  \mathrm{change}_i(S) & = & \{ p \in P~|~\exists p_S \in S~\mathrm{s.t.~move}~M_i~\mathrm{can~be~made~from}~p_S~\mathrm{to~reach~position}~p \} \\
\end{eqnarray*}

\noindent
We can then implement $\mathrm{forward}(S)$ as follows -

\begin{eqnarray*}
  \mathrm{forward}(S) & = & \bigcup_i \left( \mathrm{change}_i(S \cap \mathrm{Pre}_i) \cap \mathrm{Post}_i \right) \\
\end{eqnarray*}

\noindent
Reverse move generation can be implemented similarly by swapping the pre- and post-conditions, and using $\mathrm{change}^{-1}$ to denote the same change operation swapping before and after values.

\begin{eqnarray*}
  \mathrm{reverse}(S) & = & \bigcup_i \left( \mathrm{change}^{-1}_i(S \cap \mathrm{Post}_i) \cap \mathrm{Pre_i} \right) \\
\end{eqnarray*}

\noindent
These implementations of $\mathrm{forward}$ and $\mathrm{reverse}$ simplify the requirements for our set representations to basic set operations -- set union ($\cup$), set intersection ($\cap$), and set difference ($\setminus$) -- and the new change functions.
Whatever set representation that we choose, we will want it to be compact and fast for these operations.

\begin{figure}[ht]
  \begin{subfigure}[b]{0.5\linewidth}
    \begin{center}
      \includegraphics[width=\linewidth]{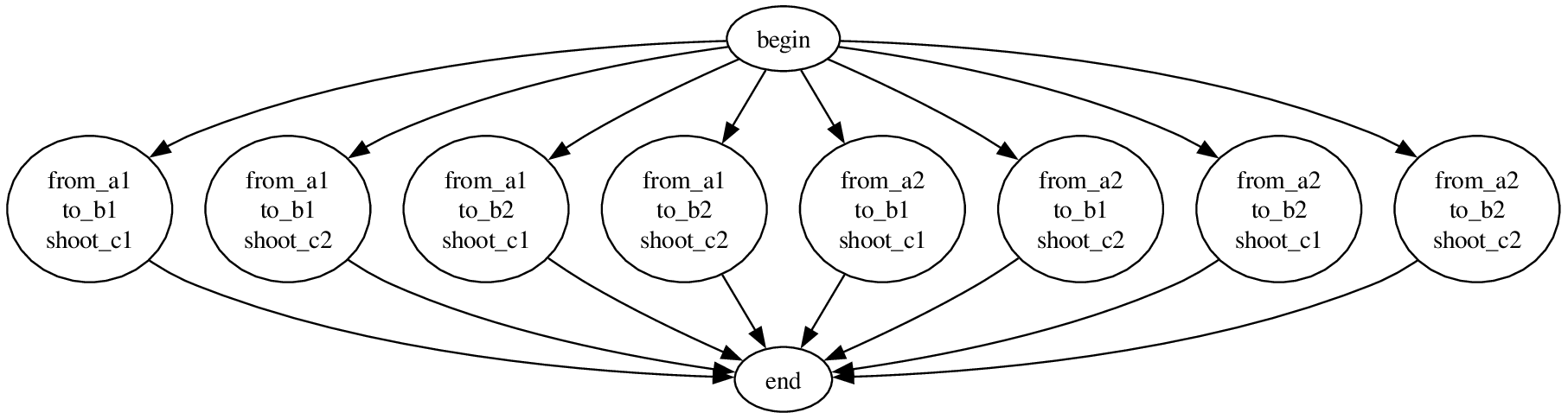}
      \caption{Simple Parallel Move Generation}
      \label{figure:move-generation-simple}
    \end{center}
  \end{subfigure}
  \begin{subfigure}[b]{0.5\linewidth}
    \begin{center}
      \includegraphics[width=\linewidth]{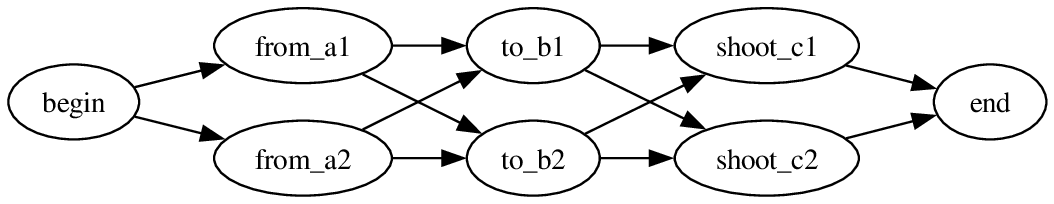}
    \end{center}

    \caption{General Move Graphs}
    \label{figure:move-generation-graph}
  \end{subfigure}

  \caption{Move Generation using Sets of Strings}
  \label{figure:move-generation}
\end{figure}

In practice, we found that this formulation of moves was theoretically sufficient, but stifling for games of only moderate complexity.
For example, if each side has more than one kind of piece, capture moves such as in Checkers or Chess need to be replicated for each type of piece captured, since the $\mathrm{Before}_{i,j}$ values differ.
Similarly, shared state changes such as clearing en-passant status in Chess would multiply all moves to handle each possible position of an en-passant pawn and the case where there was none.
To reduce these blowups in the numbers of moves, we implemented a more general move graph.
We omit the details for space, but illustrate the simple and general graph formulations in Figure~\ref{figure:move-generation}.

\subsection{Compressing Sets of Positions using Deterministic Finite Automata}

We use deterministic finite automata (DFAs) to represent sets of positions.
DFAs are a classic data structure used both in complexity theory and for parsing.
A DFA has a fixed number of states, and a transition table mapping pairs of current state and input character to the next state.
Using the transition table, a DFA can process a string in time linear in the number of characters.
Set operations with DFAs usually take quadratic time -- computing the set union, intersection, or difference of an $m$ state DFA and an $n$ state DFA can be done in $\Theta(mn)$ time.
This quadratic performance is well above our linear target, but we empirically find that the performance is closer to linear for the sets that arise solving games.
The last kind of operation that we need from our compressed set representation is the change operations.

\begin{figure}
  \begin{subfigure}[b]{0.50\linewidth}
    \begin{center}
      \includegraphics[width=0.9\linewidth]{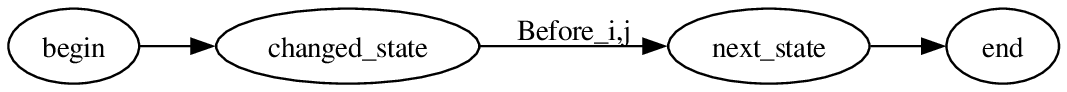}
    \end{center}
    \caption{A DFA before the change.}
  \end{subfigure}
  \begin{subfigure}[b]{0.50\linewidth}
    \begin{center}
      \includegraphics[width=0.9\linewidth]{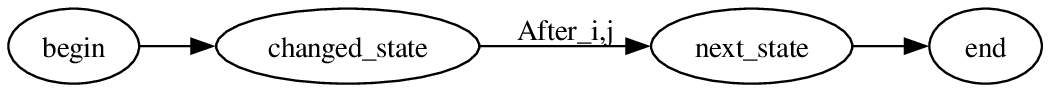}
    \end{center}
    \caption{A DFA after the change.}
  \end{subfigure}

  \caption{An example of the change operation on DFAs.}
  \label{figure:dfa-change}
\end{figure}

We can implement the change operation in time linear in the number of DFA states by rewriting the transition table in one pass.
Since we are using DFAs to encode fixed length strings with one character per board position, we can easily map individual states in the transition table to string indexes using depth-first search.
For any string index where the values will be changed, we already required the DFA to require the before value.
That means that for any DFA state for that string index, there will be at most one non-rejecting transition, and it must be for the before character.
To implement the change operation, that non-rejecting transitioned is simply swapped from the before character to the after character.
This can be done for all string indexes in one pass, so the change operations take time linear in the number of DFA states.
Additionally, change operations do not change the number of DFA states.
Figure~\ref{figure:dfa-change} shows an example DFA change.

How long do the forward and reverse functions take?
Each function starts by intersecting the input set with the pre or post condition.
Those conditions are fixed for a given move, so those take linear time and increase the number of DFA states by at most a constant factor.
Then a change operation is applied which also takes linear time and does not increase the number of states.
And then the post- or pre-condition is applied which again takes linear time and increases the number of DFA states by at most a constant factor.
So generating the results of a specific move takes linear time.
What about the final union combining all of the moves?

We have not proven bounds on the final union of forward and reverse move generation.
Empirically, the cost appears linear from our game solving usage, but we speculate that if it is linear for a given game, the constant factor is exponential in either the number of moves or the number of changes across moves.
This linear cost and constant factor blowup does mean that the total work grows exponentially with the number of ply analyzed.
This is not surprising, and we can still gain an advantage as long as the constant factor blowup tends to be lower than the branching factor of the game or relevant sets of positions.
However, it does mean that games where those factors are close will show minimal gains from our techniques.

\section{Results}
\label{section:results}

We implemented compressed game solving using deterministic finite automata, and implemented the rules of a number of different games.
We briefly check the solutions for Nim to confirm that they match the known analytical results.
We then present our headline results for Breakthrough~\cite{handsomb2001game} where we solved several new board sizes on just a laptop.
We then close with analysis of Amazons and Chess where we saw less leverage from our techniques.
All results presented here were achieved on a 2016 MacBook Pro with a 2.9GHz Quad-Core Intel Core i7, 16GB RAM and 1TB storage.

\subsection{Nim}

Nim is an ancient game with the distinction of having had a complete mathematical solution since 1901~\cite{bouton1901nim}.
It is sometimes called ``the stick game'' by children.
The game starts with a few heaps of sticks, and each player in turn picks a non-empty heap and removes at least one stick from that heap.
Depending on the version played, the goal of the game is to pick up the last stick, or avoid picking up the last stick.
We consider the normal play version of Nim where the last player to pick up the last stick wins.
For this version of Nim, a position is losing if and only if the binary XOR of all heap sizes is zero.
With such a trivial analytical solution, Nim is rarely a target of search or brute-force methods.
However, it has been noted that because the analytical solution is based on parity, deep-learning methods such as those of AlphaZero have trouble with these games~\cite{zhou2022impartial}.
We applied our methods briefly to Nim and confirmed that they match the analytical solution.
Nim is likely the most extreme case of leverage for our methods -- for an initial position of $m$ heaps of $n$ sticks each, the number of possible states is $(n+1)^m$, but the number of DFA states is only $\Theta(mn)$.

\subsection{Breakthrough}

Breakthrough is a relatively new game which won the first 8x8 Game Design Competition in 2001~\cite{handsomb2001game}.
Breakthrough has just one piece type for each player, trivial setup (two rows of pawns starting on their respective sides), and simple movement rules.
Pawns may move straight forward or diagonally forward, and may capture if moving diagonally.
A player wins if they move one of their pawns to the farthest row from their start or capture their opponent's last piece.
Those rules are sufficient to play Breakthrough, but play exhibits non-trivial complexity, as an enemy's piece may only stopped from advancing by capturing it; simply blocking it is impossible.
Previous work weakly solved Breakthrough for sizes up to 3x7, 4x6, and 6x5 using a mixture using novel search strategies, proof number search and identifying patterns~\cite{skowronski2009automated,finnsson2011gametree,saffidine2011solving}.
We are not even the first to apply retrograde analysis to Breakthrough~\cite{isaac2016generating,devink2022solving}, but anecdotally, piece-limited end-game databases do not help much, as Breakthrough games tend to end with many pieces on the board.

\begin{figure}
  \begin{subfigure}[b]{0.33\linewidth}
    \includegraphics[width=\linewidth]{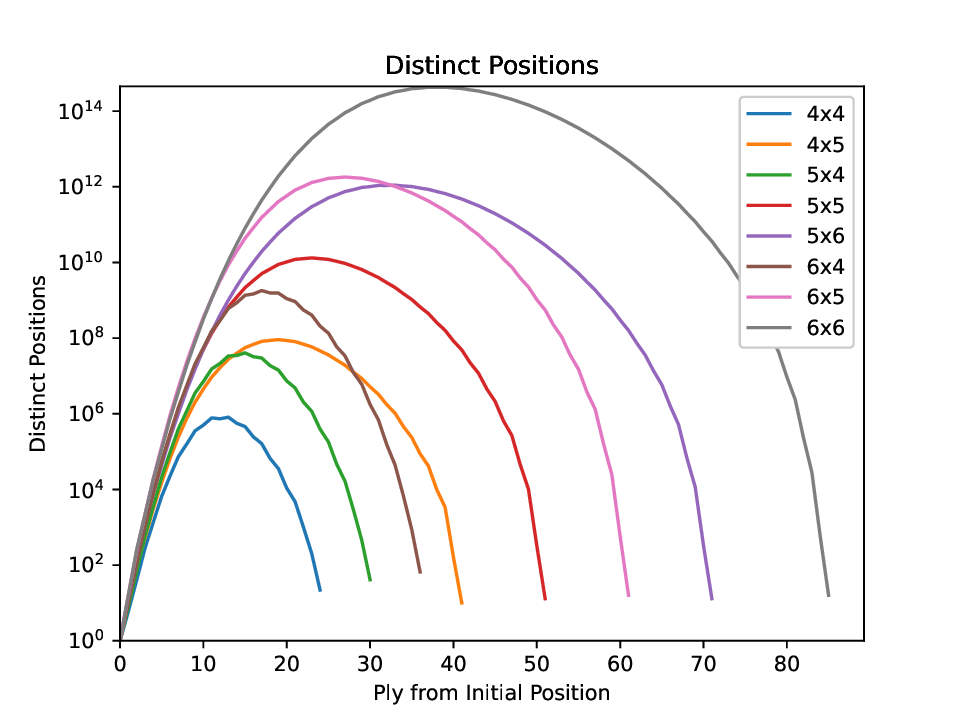}
    \caption{Breakthrough Distinct Positions}
    \label{figure:breakthrough-reachable-positions}
  \end{subfigure}
  \begin{subfigure}[b]{0.33\linewidth}
    \includegraphics[width=\linewidth]{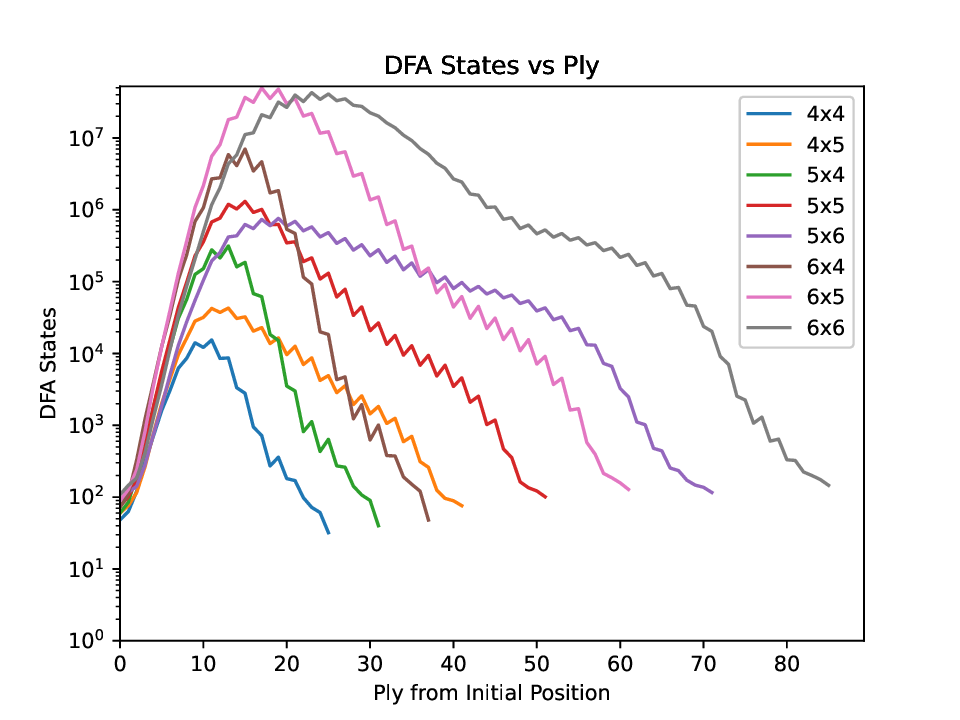}
    \caption{Breakthrough DFA States}
    \label{figure:breakthrough-reachable-states}
  \end{subfigure}
  \begin{subfigure}[b]{0.33\linewidth}
    \includegraphics[width=\linewidth]{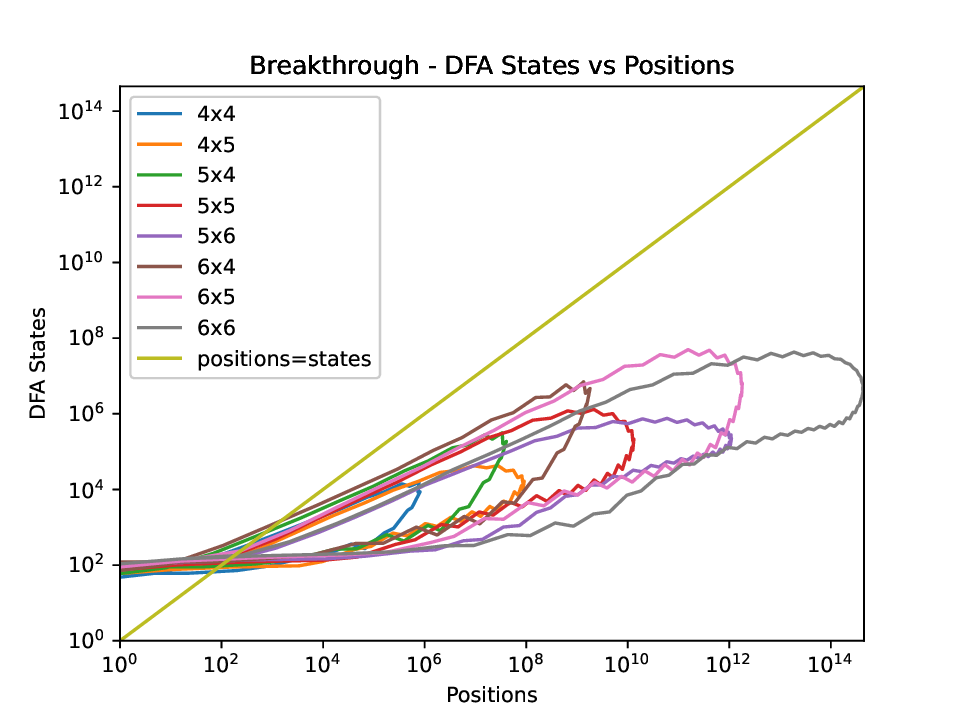}
    \caption{Breakthrough States vs Positions}
    \label{figure:breakthrough-reachable-states-vs-positions}
  \end{subfigure}

  \caption{Reachable Breakthrough Positions}
  \label{figure:breakthrough-reachable}
\end{figure}

We initially tested our Breakthrough implementation generating all the reachable positions by ply for several different board sizes.
Breakthrough games have bounded length that can be easily computed from the board size, so we were able to completely enumerate all reachable Breakthrough positions for those board sizes.
Figure~\ref{figure:breakthrough-reachable} shows those statistics.
Using the reachable set with the highest number of states, we estimated the fractional polynomial exponent at $0.45$, so $n$ positions would be represented by approximately $n^{0.45}$ states.
This gave us confidence that we had an effective compressed representation for Breakthrough.
We then applied the set-based meet-in-the-middle strategy previously described using $W_0$ and $L_0$ for our initial retrograde solutions.
We strongly solved all the sizes of breakthrough that were previously solved weakly, and additionally solved a few new bigger sizes, including the 5x6 size that thwarted previous efforts.
Table~\ref{table:breakthrough-results} shows all known breakthrough solving results to date.

\begin{table}[ht]
  \begin{center}
    \begin{tabular}{|l||l|l|l|l|l|}
      \hline
      width x height & 4 & 5 & 6 & 7 & 8 \\
      \hline
      \hline
      2 & P2 & P2 & P1 & P2 & P1 (new) \\
      \hline
      3 & P2 & P2 & P1 & P2~\cite{haugland_breakthrough3x7} & P1 (new) \\
      \hline
      4 & P2 & P2~\cite{saffidine2011solving} & P1~\cite{devink2022solving} & P1 (new) & \\
      \hline
      5 & P2 & P2~\cite{haugland_breakthrough5x5} & P1~(new) & & \\
      \hline
      6 & P2 & P2~\cite{saffidine2011solving} & & & \\
      \hline
      7 & P2 (new) & & & & \\
      \hline
      8 & P2 (new) & & & & \\
      \hline
    \end{tabular}

    \hfill \\
    P1 denotes a first player win.
    P2 denotes a second player win.
    If no citation is provided, we were unable to find a previous mention of solving this board size, but larger board sizes were solved prior to this work.
  \end{center}
  \caption{Breakthrough Solved Board Sizes}
  \label{table:breakthrough-results}
\end{table}

\begin{figure}
  \begin{subfigure}[b]{0.33\linewidth}
    \includegraphics[width=\linewidth]{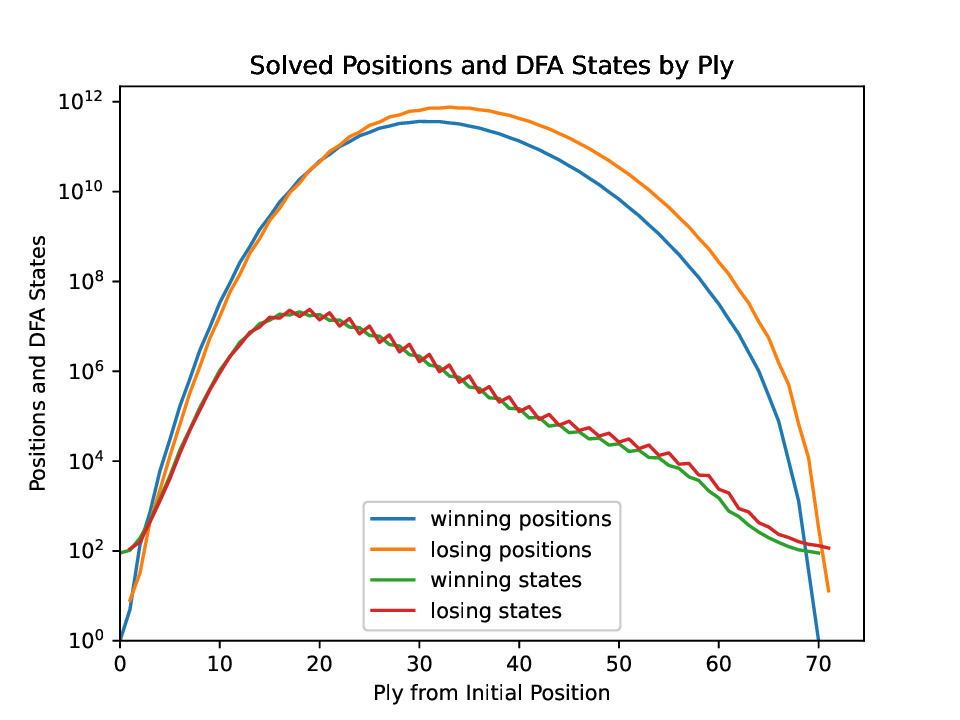}
    \caption{5x6 Solved Positions}
    \label{figure:breakthrough_5x6_solved}
  \end{subfigure}
  \begin{subfigure}[b]{0.33\linewidth}
    \includegraphics[width=\linewidth]{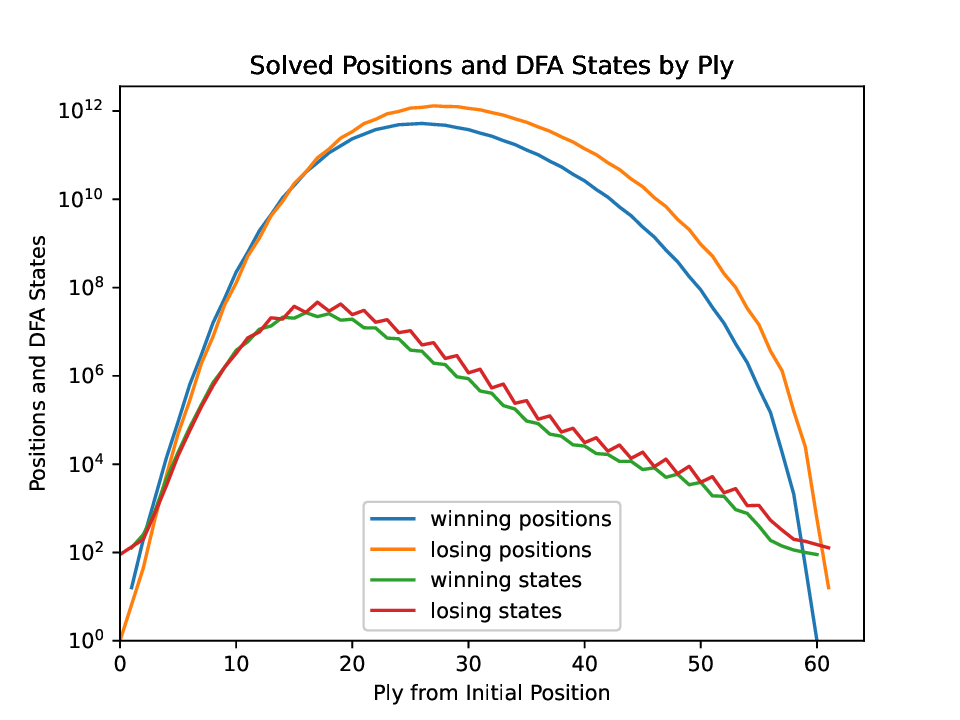}
    \caption{6x5 Solved Positions}
    \label{figure:breakthrough_6x5_solved}
  \end{subfigure}
  \begin{subfigure}[b]{0.33\linewidth}
    \includegraphics[width=\linewidth]{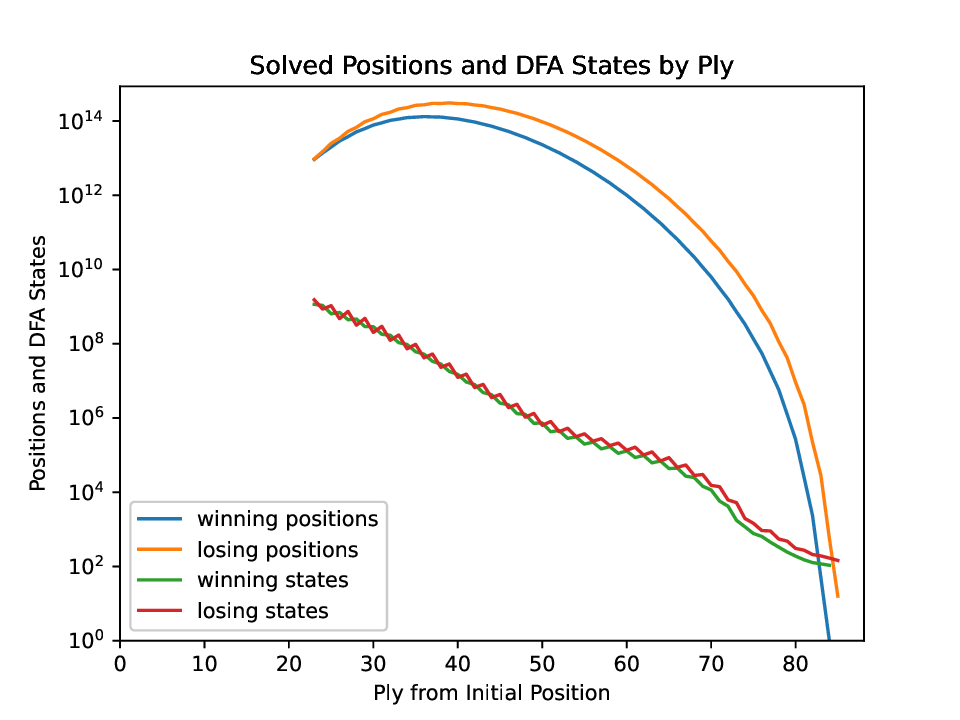}
    \caption{6x6 Solved Positions}
  \end{subfigure}

  \caption{Breakthrough Largest Solutions}
  \label{figure:breakthrough-biggest}
\end{figure}

Figure~\ref{figure:breakthrough-biggest} shows statistics for the 5x6, 6x5, and 6x6 board sizes.
The 6x5 board size was previously the largest solved size, while 5x6 previously resisted efforts.
We solved both of those easily, and saw substantial compression gains solving both -- see Figures~\ref{figure:breakthrough_5x6_solved}~and~\ref{figure:breakthrough_6x5_solved} to compare the numbers of positions solved and the states to represent them, and note those charts are in log scale.

At the time of submission, our solution to the 6x6 board size is still in progress.
We have already solved $R_{23}$ through the end of the game, and have already passed the ply with peak positions.
Separately, we have computed $W_{16}$ and $L_{16}$ for this board size and determined that for $U_{16} = \mathrm{inverse}(W_{16} \cup L_{16})$, $|U_{16}| \approx 1.3 \times 10^{11}$, so solving the remaining positions is easily within the range of brute force (and may even be faster without compression).
Even in its incomplete state, this partial solution already includes more than $7 \times 10^{15}$ reachable positions, making this the largest solution known to us by about a factor of two.
While the margin is small, this record is more striking considering that the previous largest solution used a super computer~\cite{irving2014pentago} and ours used a commodity laptop.
So far, this 6x6 solution has averaged $4.9 \times 10^4$ positions per byte with our simple DFA encoding using 12 bytes per state.
The peak compression rate has been $1.2 \times 10^7$ positions per byte to store the $9.5 \times 10^{13}$ positions of $RL_{50,\infty}$ in $6.4 \times 10^{5}$ states.
For comparison, the ten piece endgame database for Checkers averaged 154 positions per byte~\cite{schaeffer2007checkers}.

\section{Conclusions and Future Work}
\label{section:conclusions}

In this work, we argued for the used of compressed sets for game solving.
We demonstrated the efficacy of this approach with the game of Breakthrough, and look forward to solving more games using this approach.
We have already started testing these ideas with Amazons and Chess, and while we saw less leverage, they also appear to have sublinear compressed set representations.
We also intend to investigate the integration of knowledge into this approach to take advantage of symmetries and combinatorial game theory.

\bibliography{biblio}
\bibliographystyle{plain}

\end{document}